\documentclass[pra,aps,showpacs,superscriptaddress,twocolumn]{revtex4-1}
\usepackage[utf8]{inputenc}
\usepackage{graphicx}
\usepackage{amsmath}
\usepackage{amsfonts}
\usepackage{amssymb}
\usepackage{epsfig}
\usepackage{color}

\begin{document}

\newcommand{\abs}[1]{\left\vert#1\right\vert}
\newcommand{\set}[1]{\left\{#1\right\}}
\newcommand{\bra}[1]{\left\langle#1\right\vert}
\newcommand{\ket}[1]{\left\vert#1\right\rangle}
\newcommand\braket[2]{\left.\left\langle#1\right|#2\right\rangle}
\newcommand\ketbra[2]{\left.\rangle\left#1\right|#2\right\langle}
\def\I {{\rm 1} \hspace{-1.1mm} {\rm I} \hspace{0.5mm}}
\newcommand{\rosso}[1]{\color[rgb]{0.6,0,0} #1}

\title{Quantum Zeno effect and non-Markovianity in a three-level system}

\author{Antti Karlsson}
\email{antti.karlsson@utu.fi}
\affiliation{Turku Centre for Quantum Physics, Department of Physics and Astronomy, University of
Turku, FI-20014 Turun yliopisto, Finland}

\author{Francesco Francica}
\affiliation{Dip. Fisica, Universit\'a della Calabria, 87036
Arcavacata di Rende (CS) Italy}

\author{Jyrki Piilo}
\affiliation{Turku Centre for Quantum Physics, Department of Physics and Astronomy, University of
Turku, FI-20014 Turun yliopisto, Finland}

\author{Francesco Plastina}
\affiliation{Dip. Fisica, Universit\'a della Calabria, 87036
Arcavacata di Rende (CS) Italy} \affiliation{INFN - Gruppo
collegato di Cosenza, 87036 Arcavacata di Rende (CS) Italy}

\date{\today}

\begin{abstract}
We study the coexistence of the quantum Zeno effect and
non-Markovianity for a system decaying in a structured bosonic
environment and subject to a control field. The interaction with
the environment induces decay from the excited to the ground
level, which, in turn, is  coherently coupled to another
meta-stable state. The control of the strength of the coherent
coupling between the stable levels allows the engineering of both
the dissipation and of the memory effects, without modifying
neither the system-reservoir interaction, nor environmental
properties. We use this framework in two different parameter
regimes corresponding to fast (bad cavity limit) and slow
dissipation (good cavity limit) in the original, un-controlled
qubit system. Our results show a non-monotonic behavior of memory
effects when increasing the effectiveness of the Zeno-like
freezing. Moreover, we identify a new source of memory effects
which allows the persistence of non-Markovianity for long times
while the excited state has already been depleted.
\end{abstract}

\bigskip

\maketitle

\section{Introduction}

How can a flying arrow be moving, if at any instant of time when
it is observed, it is seen in some place, stationary? This was
one of the paradoxes of Zeno of Elea \citep{laertius}, an ancient
Greek philosopher. A little bit over two thousand years later, von
Neumann's reduction postulate \cite{neumann} laid the foundation
for a similar effect in quantum mechanics. Namely, if a quantum
measurement collapses the state of a quantum system to a state in
the measurement basis, this process is repeated $N$ times a second
and $N$ is let to tend to infinity, the motion of the system is
prevented. The quantum phenomenon was named the quantum Zeno
effect in \cite{sudarmisra} and has been been studied in several
earlier works,
see~\cite{beskow,khalfin, kurizki1, kurizki2, zenoreview,zenoent1,zenoent2} and
references therein. Besides frequent measurements, a strong
coupling to an external, control system or level can also prevent
the original system of interest from evolving in time
\cite{kraus,watchdog}. Loosely speaking the control system is
continuously measuring or gazing at the system of interest and
thereby preventing its dynamics. This effect is called dominated
evolution or the watchdog effect and is the one that we will study
in this paper from the point of view of non-Markovianity and
information flow.

Markovian dynamics is generally understood to be describable by
the semigroup evolution and  Lindblad equation \cite{BREUERBOOK}.
However, Markovian dynamics is always an approximation and not
necessarily valid for all systems nor environments. Therefore,
understanding non-Markovian memory effects is an important aspect
when studying open system dynamics in general. Different
approaches to define and quantify non-Markovian dynamics based on
different physical or mathematical quantities have been actively
developed in recent years
\cite{wolf,NMprl,rivas,mani,determinant,LFS}, see also recent
reviews \cite{NMreviewblp, NMreviewhuelga}. These quantifiers are
based on the non-monotonicity of some kind of information flow
towards the system, and have been compared and classified in
\cite{mani,com,comp1,compa2,compar3,compare4,compared5}.

Here, we are interested in describing the modifications of the
information flow due to the control field induced freezing of the
decay, and, more generally in how to control the Zeno effect and
non-Markovianity, and what is their interplay. For this purpose,
we study a two-level system interacting with a zero-temperature
bosonic environment. In this qubit system, the coupling strength to
the environment and environmental properties, such as spectral
density, define whether the qubit dynamics displays memory effects
or not. However, adding a coherent coupling to an auxiliary third
level allows to control the excited state dynamics -- displaying
Zeno-effect-- and also to engineer the memory effects. The paper
is organized in the following way. Section~\ref{dyna} describes
the system under study and Sec.~\ref{Nonmark} introduces briefly
the basic aspects of the used measure for non-Markovianity. The
central results are presented in Sec.~\ref{res} and
Sec.~\ref{disc} concludes.

\section{Model and Dynamics}\label{dyna}

\begin{figure}[t!]
\includegraphics[width=0.3\textwidth]{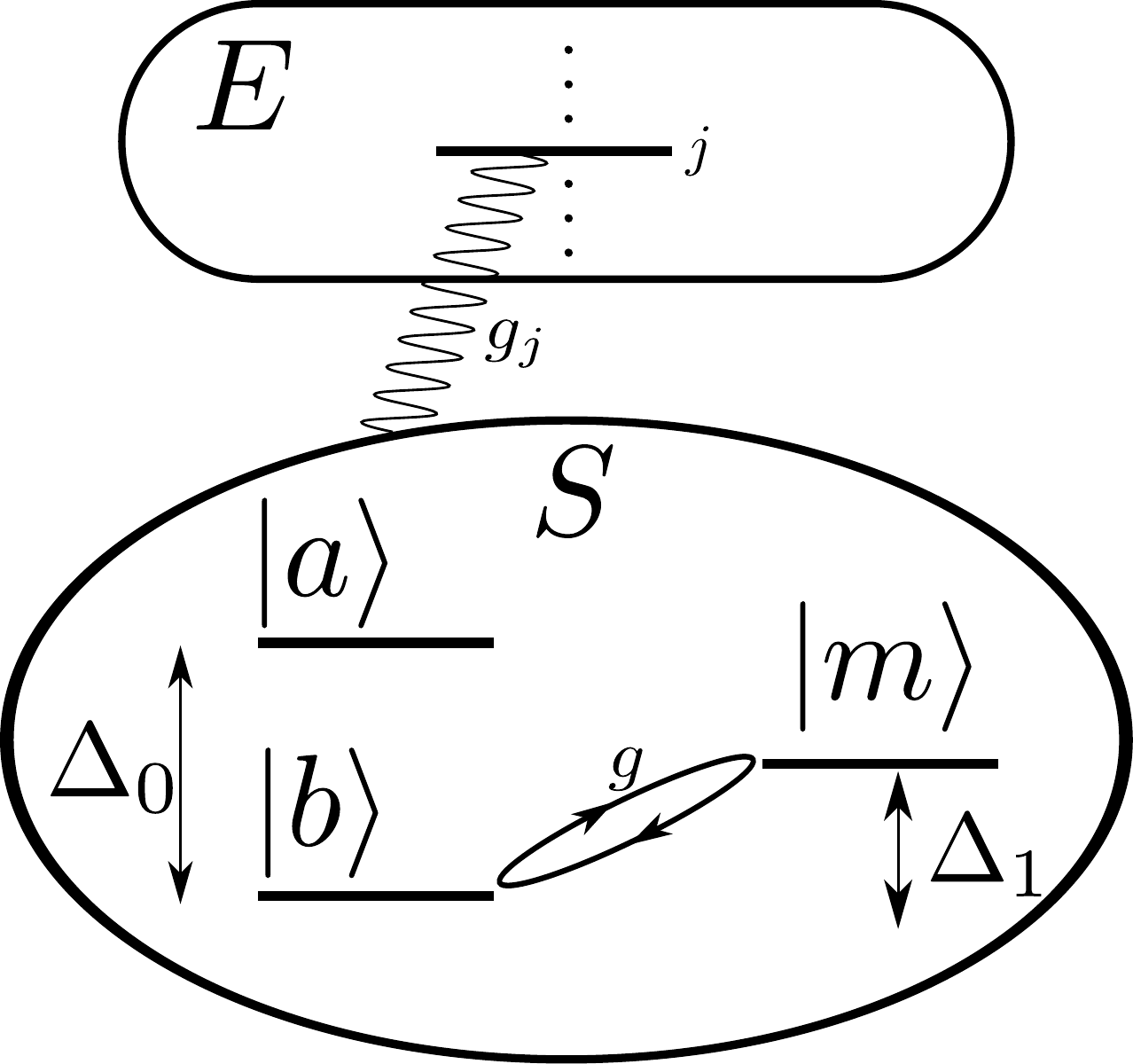}
\caption{Sketch of our model: A two-level system, with excited and
ground states $|a\rangle$ and $|b\rangle$, respectively, interacts
with a zero-temperature bosonic environment E, while the lower level
$|b\rangle$ is coherently coupled to an external level $|m\rangle$
with strength $g$. Such a coupling, enables the control of the
$|a\rangle \longrightarrow |b \rangle$ decay and of memory
effects.}\label{system}
\end{figure}

Our system is schematically described in Fig~\ref{system}. The total Hamiltonian of the system and the environment in the rotating wave approximation can be written as
\begin{align}
H_{tot} = H_S + H_E + H_{int} + H_C
\end{align}
where
\begin{align}
H_S &= \omega_a \ket{a} \bra{a} + \omega_b \ket{b} \bra{b}+\omega_m \ket{m} \bra{m} \\
H_E &= \sum_j \omega_j a_j^\dagger a_j \\
H_{int} &= \sum_j g^*_j \ket{b}\bra{a}a^\dagger_j + g_j
\ket{a}\bra{b}a_j, \\
H_C &= g (\ket{b} \bra{m} e^{i \Delta_1 t} + \ket{m} \bra{b} e^{-
i \Delta_1 t})
\end{align}
where $a^\dagger$ and $a$ are the bosonic creation and annihilation
operators, $g_j$ and $g$ the coupling strength to the $j$th mode
of the environment and the level $\ket{m}$ respectively,
$\omega_a, \hspace{2pt} \omega_b, \hspace{2pt} \omega_m$ the
frequencies of the levels $| a \rangle, | b\rangle, | m\rangle$,
$\omega_j$ are the frequencies of the environment modes and
\begin{align}
\Delta_0 &= \omega_a - \omega_b \nonumber \\
\Delta_1 &= \omega_m - \omega_b.
\end{align}
Notice that the level $| m\rangle$ is neither directly coupled
with the environment nor with level $|a\rangle$; it enters the
dynamics for control purposes only. From here on, we will work in
the interaction picture defined by the free Hamiltonian. The
interaction Hamiltonian becomes
\begin{align}
H^{(i)} = &\sum_j  g_j \ket{a}\bra{b}a_j e^{-i(\omega_j - \Delta_0)t} + g^*_j \ket{b}\bra{a}a^\dagger_j e^{i(\omega_j - \Delta_0)t} \nonumber \\
&+g(\ket{b} \bra{m} + \ket{m} \bra{b}).
\end{align}

For simplicity we consider the case of only one excitation in the
whole system initially, with initially empty environment modes.
This means that the initial state can be written as
\begin{align}
\ket{\psi(0)} = (\alpha_0 \ket{a}+\beta_0 \ket{b}+\mu_0 \ket{m})\otimes \ket{\{0\}}.
\end{align}
Because the excitation number is conserved, the state at any later
time is
\begin{align}
\ket{\psi(t)} = &(\alpha(t) \ket{a}+\beta(t) \ket{b}+\mu(t) \ket{m})\otimes \ket{\{0\}} \nonumber \\
&+ \sum_j \beta_j(t) \ket{b} \otimes \ket{1_j} + \sum_j \mu_j(t) \ket{m} \otimes \ket{1_j},
\end{align}
where $ \ket{1_j} = a^\dagger_j \ket{\{0\}}$ means an excitation
in the $j$th mode in the environment. To find the form of the
coefficients we solve the Schr\"{o}dinger equation in the
Appendix. For the bosonic environment, we use a Lorentzian
spectral density function
\begin{align}\label{specdens}
J(\omega) = \Omega_0^2 \frac{\lambda}{\pi((\omega-\Delta_0)^2 + \lambda^2)},
\end{align}
where $\Omega_0^2=\frac{\lambda \gamma}{2}$. Varying the width of
the Lorentzian allows us to use a good and a bad cavity limits --
having $\gamma / \lambda \gg 1$ ($\gamma / \lambda \ll 1$)
corresponds to good (bad) cavity limit. The solution of the
Schr\"{o}dinger equation is fairly complicated but we can in quite
a straightforward manner use it to study the dynamics of the
system numerically. In the following sections, we first introduce
the BLP-measure for quantifying the non-Markovianity and then
present the results.

\section{Non-Markovianity}\label{Nonmark}


In general, quantum channels are one-parameter families
$\{\Phi_t\}_{t>0}$ of completely positive, trace preserving (CPTP)
maps. Each member $\Phi_t$ of the family evolves a quantum state
from the initial time $t=0$ to time $t>0$, denoted
\begin{align}
\rho(t) = \Phi_t \rho(0).
\end{align}
The set of quantum states is the set of positive operators with
unit trace and can be endowed with a metric called the trace
distance $D(\cdot,\cdot)$ induced by the trace norm $||\cdot||_1$
by the following formula
\begin{align}
D(\rho_1, \rho_2) = \frac12 || \rho_1 - \rho_2 ||_1.
\end{align}
It turns out \cite{fuchsvdg} that this metric is related to the
optimal probability $P_{max}$ of correctly distinguishing two
unknown quantum states from each other. The relation is
\begin{align}
P_{max}(\rho_1,\rho_2) = \frac{1}{2}(1+D(\rho_1,\rho_2)).
\end{align}
This gives an operational meaning to the trace distance as a
measure of  distinguishability of quantum states. CPTP maps are
contractions for the trace distance \cite{ruskai}, which means
that quantum channels tend to decrease the distuinguishability of
quantum states. We defined channels as families of CPTP maps that
map the input states from initial time $t=0$ to some later time
$t$. If we study the intermediate maps $\Phi_{t_1,t_2}$, where
$0<t_1<t_2$, which evolve the state from time $t_1$ to time $t_2$,
the CPTP property need not hold anymore. This means that locally
in time, the trace distance can increase, but never above the
original value at time $t=0$.

Interpreting the decrease of trace distance as information flowing
out and increase of the trace distance as a refocusing of
information onto the system, we arrive at the Breuer, Laine, Piilo
(BLP) measure of non-Markovianity \cite{NMprl} defined by
\begin{align}
\mathcal{N}_{\textrm{BLP}} = \underset{\rho_1(0),\rho_2(0)}{\text{sup}} \int_{\sigma(\rho_1(t), \rho_2(t))>0} \sigma(\rho_1(t), \rho_2(t) )\hspace{2pt} \text{d} t,
\end{align}
where
\begin{align}
\sigma(\rho_1(t), \rho_2(t) )=\frac{d D(\rho_1(t), \rho_2(t))}{d t}.
\end{align}
The measure takes as an input a pair of initial states $\rho_1(0)$
and $\rho_2(0)$, monitors the dynamics of their trace distance and
adds up all the possible increases in time. This is maximized over
all possible choices of the initial pairs giving a value that
describes the non-Markovianity of a quantum channel.

The maximization over all possible pairs of initial states seems
cumbersome, but can be simplified \cite{optpairs,univnm}. The idea
is, that since the state pairs enter the measure only as their
difference, the important quantity is actually the direction in
the set of quantum states. Many different state pairs give the
same direction, meaning that their difference is the same up to a
constant. Also, any direction (which here means a traceless
hermitian matrix) can be written as a difference of two quantum
states \cite{univnm}. This simplifies the maximization procedure
by removing the need to search over all possible state
combinations and replacing it with a search over all possible
directions. We use this method in our numerical work, which is
presented in the following section.

\section{Results}\label{res}

Since the external level $\ket{m}$ is introduced only to control
the dynamics of the original two--level system, we restrict the
optimization to initial states where the external level
$|m\rangle$ is empty. This means that we are interested in initial
states with $\mu_0 = 0$, which in turn means that the states are
described by the parameters $\alpha_0$ and $\beta_0$ alone. This
implies that in the initial state density matrix (see the appendix), only the upper
left 2x2 block is non-zero, which enables us to represent the
interesting ones using the Bloch sphere, i.e in the form
\begin{align}
\rho_0 = \frac{1}{2}( \sigma_0 + \vec{r} \cdot \vec{\sigma}),
\end{align}
where $||\vec{r}||\leq 1$ and $\sigma_i$ are the Pauli matrices.
The matrix describing the actual state of our system is the form
$\rho_0$ above, appropriately padded with zeros to make the
density matrix 3x3. We find the value of the measure by exploiting
the direction argument in the following way. First we choose a
finite integration time, $\lambda t=20$, for which the evolution
of the system is monitored.  A random point and its antipodal
point from the Bloch sphere are chosen as the state pair for which
the value of the trace distance integral is calculated f. This
number is stored and the process repeated for a desired number of
times (in our case 500 samples). The numbers are normalized to
range from 0 to 1. This way we get an approximate map of how the
different directions on the Bloch sphere behave in terms of the
BLP measure.

\begin{figure}
\includegraphics[width=0.3\textwidth]{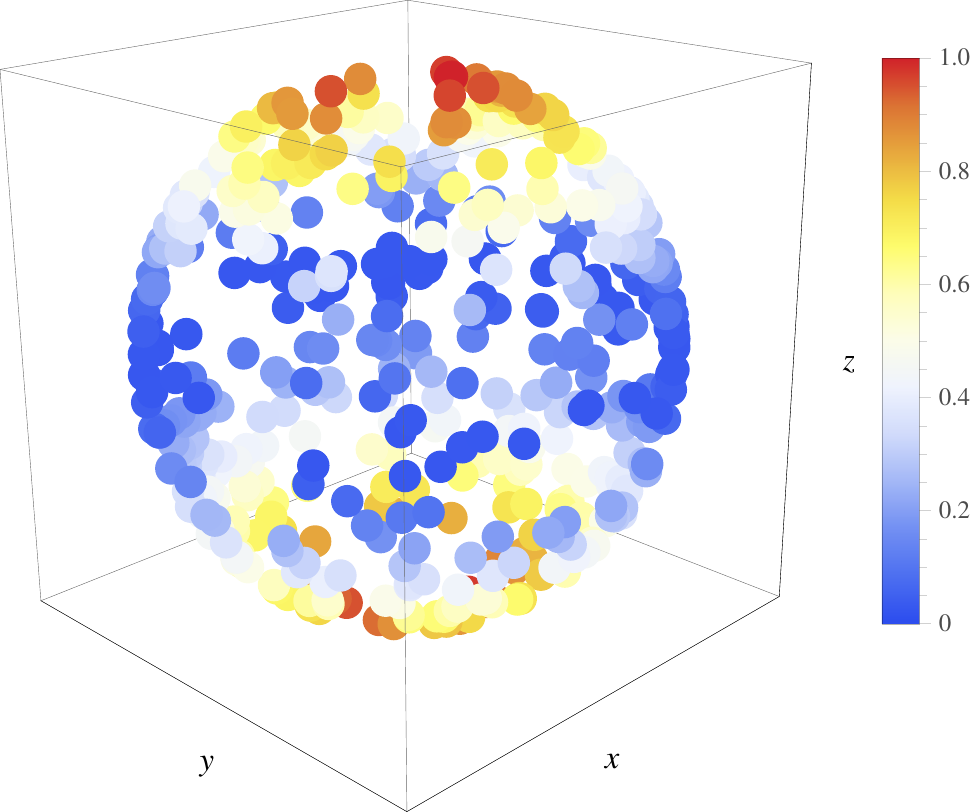}
\caption{(Color on-line): Bloch sphere representation of the
directions and the corresponding value for the BLP measure with
parameter values $\gamma=10\lambda, \hspace{2pt} g=10 \lambda$.
Red means high and blue low value for the measure. Note that in
the figure, the value of the measure has been normalized so that
$0\leqslant \mathcal{N}_{\textrm{BLP}}\leqslant
1$.}\label{spherepic}
\end{figure}

We study two different regimes -- corresponding to a bad and a
good cavity -- by appropriately choosing the parameter $\gamma$ of
the spectral density defined in equation~\eqref{specdens}. In
particular, we take $\gamma=\frac{\lambda}{10}$ to describe a bad
cavity and $\gamma=10 \lambda$ for a good cavity. Also the
coupling strength $g$ to the external level is varied between 0
and $100 \lambda$. In all but one of the cases that we studied,
the values of the measure were distributed like in Fig.
\ref{spherepic}, meaning that the maximizing direction is the one
through the north and the south pole, corresponding to the initial
state pair
\begin{align}
\rho_1(0) &=  |a\rangle \langle a |  \\
\rho_2(0) &=  |b\rangle \langle b |.
\end{align}
In the one exception, which was the good cavity with zero coupling
to the external level, the maximizing pair could have been taken
as any pair from the equator.

\begin{figure}[t]
\includegraphics[width=0.5\textwidth]{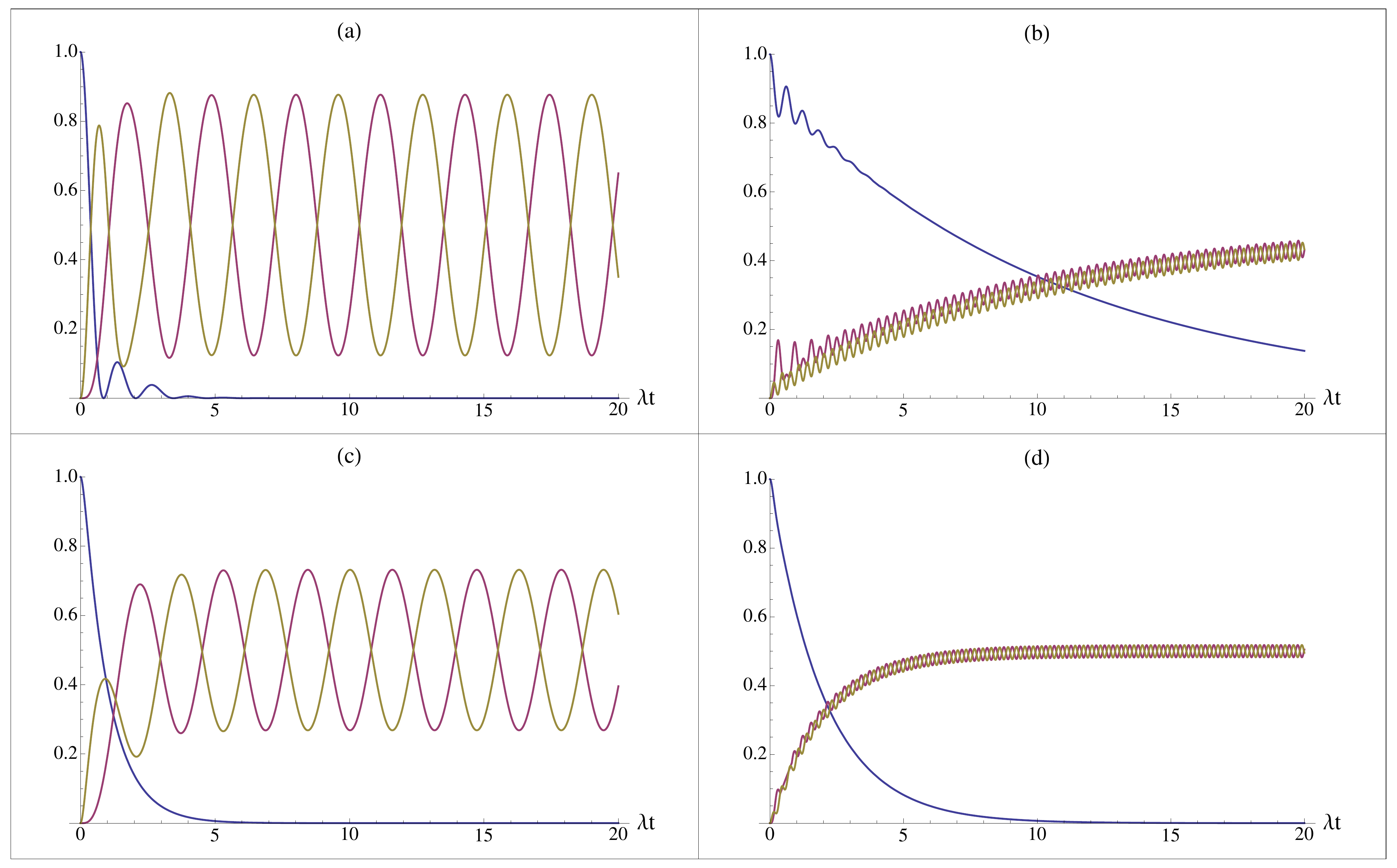}
\caption{(Color on-line): Population of the excited state (blue),
ground state (yellow)  and the external level (purple) in the good
(a-b) ($ \gamma=10 \lambda$) and bad (c-d) cavity
($\gamma=\frac{\lambda}{10}$) case. In (a) we have used
$g=\lambda$ and in (b) $g=10\lambda$. Similarly in (c)
$g=\frac{\lambda}{10}$ and in (d) $g=\lambda$. In all of these
cases, we took an initially excited system prepared in $|a
\rangle$.} \label{alphas}
\end{figure}

In Figure~\ref{alphas}, we plot the dynamics of the population of
the different levels to show how increasing the coupling $g$
affects the dynamics of the system both in good and bad cavity
cases. In Figure~\ref{alphas} (a), which corresponds to good
cavity with weak coherent coupling between the two lower levels,
the excited level population reaches zero followed by a few
revival cycles typical for non-Markovian behaviour. The
populations of the lower levels keep oscillating with quite a
large amplitude and small frequency. When the coupling $g$ is
increased, Fig.~\ref{alphas} (b), the Zeno effect influences the
dynamics making the dissipation from the excited state slower.
This also decreases the oscillation amplitudes of the populations
of the lower levels, and at the same time frequency is higher due
to the increased value of $g$.  For the bad cavity case in
Fig.~\ref{alphas} (c) and (d) the situation looks qualitatively
similar except that the excited state decreases monotonically in
contrast to oscillations displayed by the good cavity case. Due to
the monotonic decrease in the excited state population, one would
be tempted to conclude that for the bad cavity case the dynamics
is Markovian. However, this is not true and eventually it turns
out that memory effects influence the dynamics in quite a long
time scale even beyond the point when the excited state has
already been depleted of population.

\begin{figure}
\includegraphics[width=0.4\textwidth]{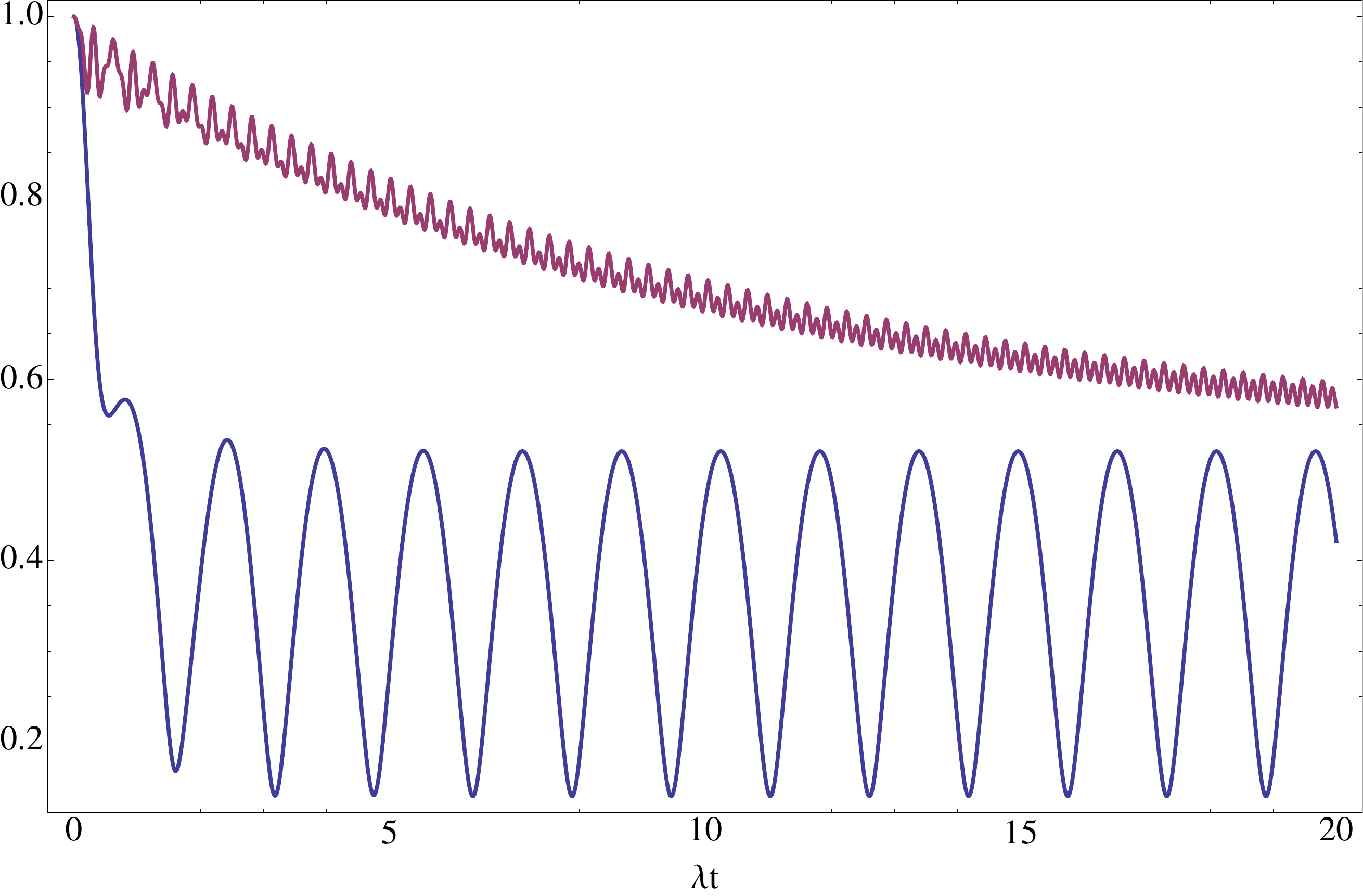}
\caption{(Color on-line): The behavior of the trace distance of
the maximizing pair in the good cavity case with $g=\lambda$
(blue) and $g=10 \lambda$ (yellow).} \label{trdist}
\end{figure}

The long term influence of the memory effects is displayed in
Fig.~ \ref{trdist}, which shows the trace distance dynamics for
the good cavity case for weak and strong coupling. For the weak
coupling, when the Zeno effect does not yet dominate the dynamics,
the trace distance keeps oscillating with quite a high amplitude
without damping beyond the point when the excited state is already
depleted. This looks peculiar since in this regime the system and
the environment do not exchange energy anymore. However, this can
be explained when looking at the equations of motions and
solutions for the various probability amplitudes (for full
details, see the Appendix, where an analytic solution of the
equations of motion is presented). First, the equation for the
excited state amplitude $\alpha$ is of the form
\begin{align}\label{alphadot}
\dot{\alpha}(t) = - \int_0^t f(t-t_1) \cos(g (t-t_1)) \alpha(t_1)
\text{d} t_1 \, .
\end{align}
By increasing $g$ the kernel of the integral keeps oscillating
faster and faster, so that the integral itself decreases, giving
rise to a freezing of the excited state amplitude. In general, for
an initial state with $\beta=\mu=0$, the open system state at time
$t$, is,
\begin{align}
\rho(t)=
\begin{pmatrix}
|\alpha(t)|^2 & 0 & 0 \\
0& \sum\limits_j |\beta_j(t)|^2 &  \sum\limits_j \beta_j(t) \mu^*_j(t) \\
0 &  \sum_j \beta^*_j(t) \mu_j(t) &  \sum\limits_j |\mu_j(t)|^2
\end{pmatrix}.
\end{align}
The time evolution of the populations of the two lower levels, also in the regime when excited state is depleted, are given by
\begin{align}
\sum_j | \mu_j (t)|^2 = \Omega_0^2 &\int_0^{t} \text{d}t_1\int_0^{t} \text{d}t_2 e^{-\lambda|t_1-t_2|} \alpha(t_1) \alpha^*(t_2) \nonumber \\
& \times \sin (g(t-t_1))\sin (g(t-t_2))  \\
\sum_j | \beta_j (t)|^2 = \Omega_0^2 &\int_0^{t} \text{d}t_1\int_0^{t} \text{d}t_2 e^{-\lambda|t_1-t_2|} \alpha(t_1) \alpha^*(t_2) \nonumber \\
& \times \cos (g(t-t_1))\cos (g(t-t_2)) .
\end{align}
The important point to notice here is that even though after some
time $\alpha=0$, the $\sin$ and $\cos$ terms depend on $t$, and
therefore also the values of the integrals depend on time and so
do the populations. This is ultimately due to the coupling between
the lower levels. However, it makes a large difference, in terms
of information flow and non-Markovianity, whether the lower levels
are coupled without prior dissipation [see Eqs.~(30-31) in the
Appendix], or can enter the coupling cycle after some population
has decayed from the excited state. As the equations above show,
due to the memory effects the lower level populations depend on
the past values of the excited state amplitude, and not only on
the instantaneous ones, thus explaining the long time survival of
oscillations in the trace distance.

The trace distance measure for non-Markovianity is generally
associated to a back-flow of information into the open system, so
that a question naturally arises: Where does the information come
from in this case? To answer, let us consider the total system
state as a function of time, for initial states with
$\beta=\mu=0$, and after the excited state has decayed,
$\alpha=0$. It is
 \begin{align}
\ket{\psi(t)} = \sum_j \beta_j(t) \ket{b} \otimes \ket{1_j} + \sum_j \mu_j(t) \ket{m} \otimes \ket{1_j}.
\end{align}
Taking a trace over the system shows that the environmental state
does not change anymore. However, the coherences within the total
system state do change and also the system-environment
correlations. It is this change of the correlations which is
ultimately responsible for long-term memory effects here. Note
that these memory effects are induced by coherently coupling the
second lower to the ground state. In other words, in this case,
the origin of the memory effects is not in the engineering of the
environment properties, nor in changing the system-environment
coupling. Instead, it is related to manipulating the global
coherences in the total system, which is enabled by the coherent
coupling.

\begin{figure}[t!]
\includegraphics[width=0.45\textwidth]{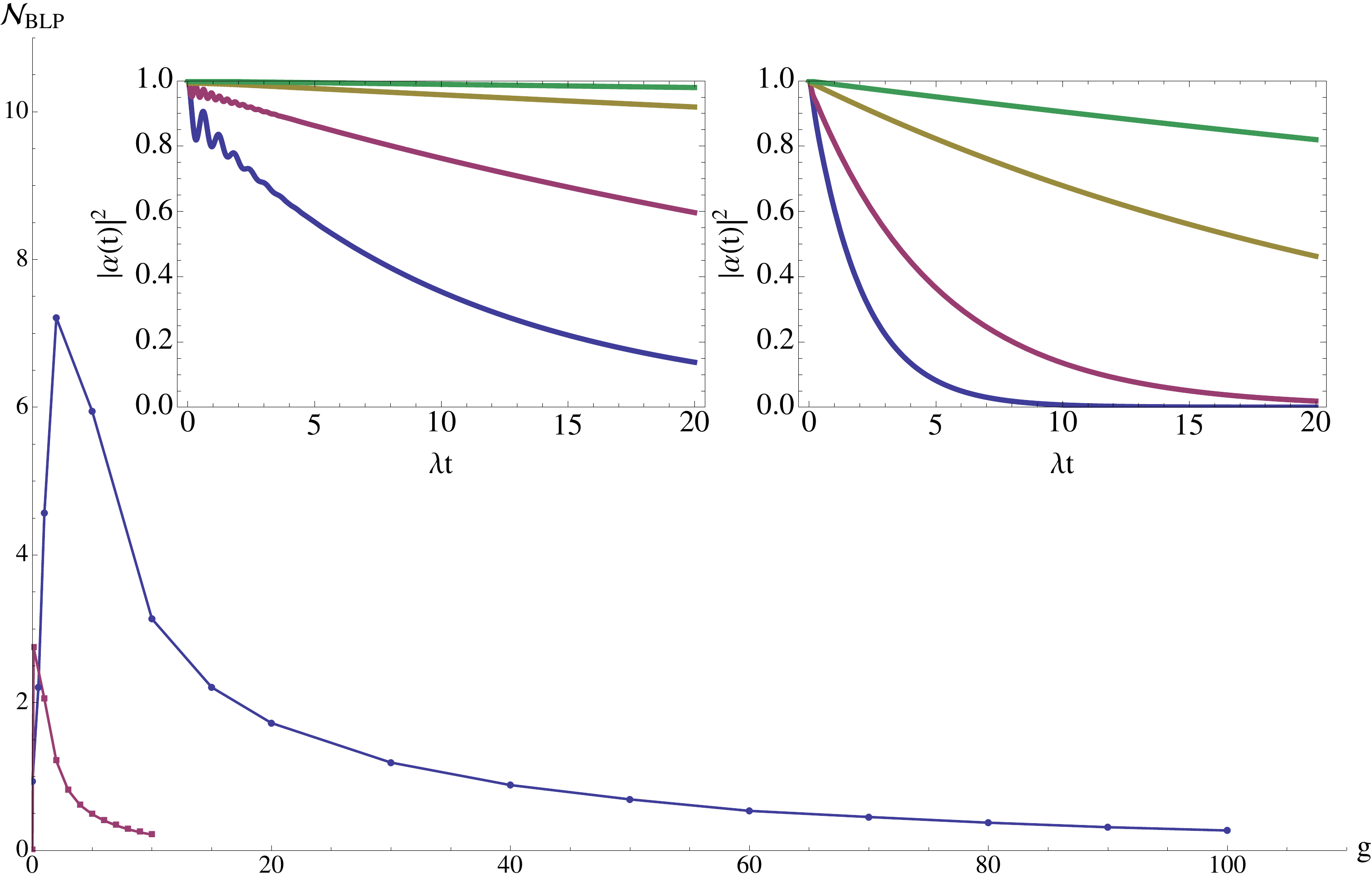}
\caption{(Color on.line): The BLP measure as a function of the
coupling strength $g$ to the external level. The blue line is for
the good cavity regime $(\gamma=10 \lambda)$, while purple refers
to the bad cavity case $(\gamma=\frac{\lambda}{10})$. The left
inset displays the population of the excited state as a function
of time with $g=10\lambda,20\lambda,50\lambda,100\lambda$ (from
bottom to top) in the good cavity case. The right inset gives the
same in the bad cavity case with
$g=\lambda,2\lambda,5\lambda,10\lambda$ (from bottom to top).}
\label{Ng}
\end{figure}

Let us now turn the attention to the amount of memory effects when
increasing the coupling $g$ and Zeno effect. The BLP measure for
the good and bad cavity cases as a function of $g$ is displayed in
Fig.~\ref{Ng}. The memory effects are more pronounced in the good
cavity case than in the bad one. However, in both of the cases the
amount of memory effects behaves in non-monotonic way. There is a
specific value of $g$ where the maximum is reached. This is
inherently related to the fact that within the current system,
there are two sources of non-Markovianity. Small-time oscillations
in the excited state population and long-time persistent
oscillations for the ground state populations. When $g=0$, the bad
cavity case does not display memory effects  whereas the good
cavity case displays minor memory effects. Increasing the coupling
constant $g$ induces the ground state oscillations with increasing
amplitude making the memory effects more prominent. However, at
the same time the Zeno-effect tends to freeze excited state
population and as a consequence, reduce the ground state
oscillations. Therefore, due to the these competing effects, a
specific value of $g$ allows to maximize the memory effects and
beyond this point Zeno-effect begins to dominate reducing
non-Markovianity. See also the insets in Fig.~\ref{Ng} displaying
in detail how Zeno-effect appears freezing the excited state
dynamics.

\section{Conclusions}\label{disc}

We have studied both Zeno-effect and non-Markovianity in a
three-level system. The results show that the same coupling,
which is used to freeze the excited state dynamics, also induces
non-Markovianity in non-trivial manner. In particular, the memory
effects persist for long times when the system and the environment
do not exchange energy anymore. As a matter of fact, there is a
competition between the strength of the memory effects and
freezing of the excited state population. As a consequence, the
amount of non-Markovianity behaves in non-monotonic way in terms
of the strength of Zeno effect. Eventually, when the population
dynamics of the excited state completely freezes, memory effects
disappear.  However, we have revealed a parameter regime which
displays a rich interplay between Zeno and non-Markovian dynamics
and this also identifies a novel source for memory effects whose
origin is inherently independent of the properties of the
environment.

{\bf Aknowledgement} We acknowledge  support  from  EU  project
QuProCS (Grant Agreement 641277) and from the Magnus Ehrnrooth foundation.

\newpage

\section*{Appendix}
Let us assume that initially there is only one excitation in the
system and that the environment modes are empty. Then the initial
state can be written as
\begin{align}
\ket{\psi(0)} = (\alpha_0 \ket{a}+\beta_0 \ket{b}+\mu_0 \ket{m})\otimes \ket{\{0\}}.
\end{align}
Since the excitation number is conserved, the state at any later
time is
\begin{align}\label{rhot}
\ket{\psi(t)} = &(\alpha(t) \ket{a}+\beta(t) \ket{b}+\mu(t) \ket{m})\otimes \ket{\{0\}} \nonumber \\
&+ \sum_j \beta_j(t) \ket{b} \otimes \ket{1_j} + \sum_j \mu_j(t) \ket{m} \otimes \ket{1_j},
\end{align}
where $ \ket{1_j} = a^\dagger_j \ket{\{0\}}$ is the state with one
excitation in the $j$th mode of the environment. Equivalently, the
state in density matrix form $\rho(t)$ after taking the partial
trace over the environmental degrees  of freedom from equation~(\ref{rhot}) is
\begin{widetext}
\begin{align}
\begin{pmatrix}
|\alpha(t)|^2 & \alpha(t) \beta^*(t) & \alpha(t) \mu^*(t) \\
\alpha^*(t) \beta(t) & |\beta(t)|^2 + \sum\limits_j |\beta_j(t)|^2 & \beta(t) \mu^*(t) + \sum\limits_j \beta_j(t) \mu^*_j(t) \\
\alpha^*(t) \mu(t) & \beta^*(t) \mu(t) + \sum_j \beta^*_j(t) \mu_j(t) & |\mu(t)|^2 + \sum\limits_j |\mu_j(t)|^2
\end{pmatrix}.
\end{align}
\end{widetext}
Schr\"{o}dinger equation now leads to the following set of coupled
differential equations for the coefficients
\begin{align}
\dot{\beta}(t) &= -i g \mu(t) \\
\dot{\mu}(t) &= -i g \beta(t) \\
\dot{\alpha}(t) &= -i \sum_j g_j e^{-i(\omega_j - \Delta_0)}\beta_j(t) \\
\dot{\beta}_j(t) &= -i g^*_j e^{i(\omega_j - \Delta_0)}\alpha(t) - i g \mu_j(t) \label{betaj} \\
\dot{\mu}_j(t) &= -i g \beta_j(t). \label{muj}
\end{align}
From the above equations we can directly solve for two coefficients
\begin{align}
\beta(t) &= \beta_0 \cos (gt) - i \mu_0 \sin (gt) \\
\mu(t) &= \mu_0 \cos (gt) - i \beta_0 \sin (gt).
\end{align}
To proceed, we use the following transformation to decouple equations~(\ref{betaj}) and~(\ref{muj})
\begin{align} \label{auxcfs}
l_j(t) = \frac{\beta_j(t) + \mu_j(t)}{\sqrt{2}} \quad \quad r_j(t) = \frac{\beta_j(t) - \mu_j(t)}{\sqrt{2}},
\end{align}
which leads to differential equations that can be integrated directly
\begin{align}
\dot{l}_j(t) &= -i \frac{g^*_j}{\sqrt{2}} e^{i(\omega_j - \Delta_0)t} \alpha(t)  - i g l_j(t) \\
\dot{r}_j(t) &= -i \frac{g^*_j}{\sqrt{2}} e^{i(\omega_j - \Delta_0)t} \alpha(t)  + i g r_j(t).
\end{align}
Integrating the above equations and solving for $\beta_j$ and $\mu_j$ from equation~(\ref{auxcfs}) yields
\begin{align}
\beta_j(t) = -ig^*_j \int_0^t \text{d}t_1 e^{i(\omega_j - \Delta_0)t_1} \cos(g (t - t_1)) \alpha(t_1) \\
\mu_j(t) = -ig^*_j \int_0^t \text{d}t_1 e^{i(\omega_j - \Delta_0)t_1} \sin(g (t - t_1)) \alpha(t_1).
\end{align}
Now we can insert the solution for $\beta_j$ to the differential equation for $\alpha$, which leads to
\begin{align}
\dot{\alpha}(t) = - \int_0^t \sum_j | g_j |^2 e^{-i (\omega_j - \Delta_0) (t-t_1)} \cos(g (t-t_1)) \alpha(t_1) \text{d} t_1.
\end{align}
To proceed from here we approximate the state of the environment
with a continuous distribution of modes, whose spectral density is
given by the function $J(\omega)$. The approximation amounts to
the replacement
\begin{align}\label{fintegral}
 \sum_j | g_j |^2 e^{-i (\omega_j - \Delta_0) (t-t_1)} \rightarrow &\int_{-\infty}^\infty J(\omega) e^{-i (\omega - \Delta_0) (t-t_1)}\text{d} \omega \nonumber \\
 &\equiv f(t-t_1).
\end{align}
With this, the equation for $\alpha$ becomes
\begin{align}
\dot{\alpha}(t) = - \int_0^t f(t-t_1) \cos(g (t-t_1)) \alpha(t_1) \text{d} t_1.
\end{align}
Let us assume that the form of the spectral density is a Lorentzian
\begin{align}
J(\omega) = \Omega_0^2 \frac{\lambda}{\pi((\omega-\Delta_0)^2 + \lambda^2)},
\end{align}
where $\Omega_0^2=\frac{\lambda \gamma}{2}$. The form of the
spectral density function could be almost anything, but this
choice makes the calculations fairly simple. By controlling the
parameters $\gamma$ and $\lambda$, which are basically the height
and width of the Lorentzian, we can switch between Markovian and
non-Markovian behavior of the system. We can now evaluate the
integral in equation~(\ref{fintegral}) and obtain
\begin{align}
f(t-t_1)= \Omega_0^2 e^{-\lambda |t-t_1|}.
\end{align}
Let us denote
\begin{align}
F(t-t_1) = \Omega_0^2 e^{-\lambda |t-t_1|} \cos(g(t-t_1))
\end{align}
and denote $t-t_1=\tau$. Then the Laplace transform of $F$ is
\begin{align}\label{lapF}
\tilde{F}(s) = \int_0^\infty e^{-s \tau} F(\tau) \text{d} \tau = \Omega_0^2 \frac{s + \lambda}{(s+\lambda)^2+g^2}.
\end{align}
Laplace transforming the differential equation for $\alpha$ we get
\begin{align}
s \tilde{\alpha}(s)  - \alpha_0 = - \tilde{F}(s) \tilde{\alpha}(s),
\end{align}
which combined with equation~(\ref{lapF}) leads to
\begin{align}
\tilde{\alpha}(s) = \alpha_0 \frac{(s+\lambda)^2 + g^2}{s(s+\lambda)^2 + s (\Omega_0^2 + g^2) + \Omega_0^2 \lambda}.
\end{align}
Denoting the three roots of the denominator with $s_i$ and taking the inverse transform, we solve for
\begin{align} \label{alphat}
\alpha(t) = &\alpha_0 \big( \frac{(s_1+\lambda)^2 + g^2}{(s_1-s_2)(s_1-s_3)}e^{s_1 t} + \frac{(s_2+\lambda)^2 + g^2}{(s_2-s_1)(s_2-s_3)}e^{s_2 t} \nonumber \\
&+ \frac{(s_3+\lambda)^2 + g^2}{(s_3-s_1)(s_3-s_2)}e^{s_3 t} \big ).
\end{align}
With this solution and after some simplifications we finally solve for the remaining coefficients in the density matrix
\begin{align}
\sum_j | \mu_j (t)|^2 = \Omega_0^2 &\int_0^{t} \text{d}t_1\int_0^{t} \text{d}t_2 e^{-\lambda|t_1-t_2|} \alpha(t_1) \alpha^*(t_2) \nonumber \\
& \times \sin (g(t-t_1))\sin (g(t-t_2))  \\
\sum_j | \beta_j (t)|^2 = \Omega_0^2 &\int_0^{t} \text{d}t_1\int_0^{t} \text{d}t_2 e^{-\lambda|t_1-t_2|} \alpha(t_1) \alpha^*(t_2) \nonumber \\
& \times \cos (g(t-t_1))\cos (g(t-t_2)) \\
\sum_j \mu^*_j(t)\beta_j(t) = \Omega_0^2 &\int_0^{t} \text{d}t_1\int_0^{t} \text{d}t_2 e^{-\lambda|t_1-t_2|} \alpha(t_1) \alpha^*(t_2) \nonumber \\
& \times \cos (g(t-t_1))\sin (g(t-t_2)). \label{mbjt}
\end{align}

\end{document}